\documentclass[superscriptaddress,twocolumn,amsmath,amssymb]{revtex4-1}
\pdfoutput=1
\usepackage{graphicx}
\usepackage{eurosym}
\usepackage{amsmath}
\usepackage{amssymb}
\usepackage{dcolumn}
\usepackage{bm}
\usepackage{subfigure}
\usepackage[final]{pdfpages}
\usepackage{soul}

\input{Qcircuit}

\def\avg#1{\mathinner{\langle{#1}\rangle}}
\def\bra#1{\mathinner{\langle{#1}|}}
\def\ket#1{\mathinner{|{#1}\rangle}}
\newcommand{\braket}[2]{\langle #1|#2\rangle}

\newcommand{\floor}[1]{\lfloor #1 \rfloor}

\newcommand{\ignore}[1]{}

\newcommand{\be}{\begin{equaArXivtion}}
\newcommand{\ee}{\end{equation}}
\newcommand{\ba}{\begin{eqnarray}}
\newcommand{\ea}{\end{eqnarray}}

\begin{document}

\title{Clock Quantum Monte Carlo: an imaginary-time method for real-time quantum dynamics}

\author{Jarrod R. McClean} 
\affiliation{Department of Chemistry and Chemical Biology, Harvard University, Cambridge MA, 02138}
\author{Al\'an Aspuru-Guzik}
\affiliation{Department of Chemistry and Chemical Biology, Harvard University, Cambridge MA, 02138}

\begin{abstract} 
In quantum information theory, there is an explicit mapping between general unitary dynamics and Hermitian ground state eigenvalue problems known as the Feynman-Kitaev Clock.  A prominent family of methods for the study of quantum ground states are quantum Monte Carlo methods, and recently the full configuration interaction quantum Monte Carlo (FCIQMC) method has demonstrated great promise for practical systems.  We combine the Feynman-Kitaev Clock with FCIQMC to formulate a new technique for the study of quantum dynamics problems.  Numerical examples using quantum circuits are provided as well as a technique to further mitigate the sign problem through time-dependent basis rotations.  Moreover, this method allows one to combine the parallelism of Monte Carlo techniques with the locality of time to yield an effective parallel-in-time simulation technique.  
\end{abstract}

\maketitle

\section{Introduction}
Understanding the evolution of quantum systems is a central problem in physics and the design of emerging quantum technologies.  However, exact simulations of quantum dynamics suffer from the so-called curse of dimensionality~\cite{Tannor:2007}.  That is, the dimension of the Hilbert space grows exponentially with the size of the physical system.  An effective remedy for the curse of dimensionality in some classical systems has been the use of Monte Carlo methods, which in many cases has an error with respect to number of samples that is independent of the dimension of the simulated system~\cite{Binder:2010}.  Unfortunately this favorable scaling is often lost in quantum systems of interest due to the emergence of the famous sign problem.  In particular, it has hindered the use of Monte Carlo methods for fermionic systems, where it is sometimes called ``the fermion sign problem'', and for real-time dynamics of general quantum systems, where it is known as ``the dynamical sign problem''.  The generic sign problem has been proven to belong to the computational complexity class NP-Complete~\cite{Troyer:2005}, and recent studies of complexity have refined knowledge about the computational power of sign-problem free (or ``stoquastic'') Hamiltonians~\cite{Bravyi:2008,Bravyi:2009}.  However, these results do not preclude the effective use of these methods on many interesting instances of physical problems.

In particular, despite the generic challenges of the sign problem, Monte Carlo methods have been used with great success in the study of electronic systems, providing a standard of accuracy in quantum chemistry and condensed matter ~\cite{Hammond:1994,Nightingale:1999,Baroni:1999,Foulkes:2001,Zhang:2003}.  In some of these methods, such as fixed node diffusion Monte Carlo, the use of a trial wavefunction allows one to approximately remove the complications of the sign problem at the cost of a small bias in the resulting energy.  One alternative to such an approximation is the use of interacting walker methods ~\cite{Zhang:1991}, which attempt to solve the problem exactly without the bias introduced by a trial function.  Recently, Booth et. al introduced an interacting walker method in the discrete basis of Slater determinants called Full Configuration Interaction Quantum Monte Carlo (FCIQMC) ~\cite{Booth:2009}. The sign problem in the context of this algorithm has been studied in some detail~\cite{Spencer:2012,Kolodrubetz:2013,Shepherd:2014} and it has been successfully applied to both small molecular systems of chemical interest and extended bulk systems~\cite{Booth:2010,Booth:2013}.  

The use of Monte Carlo methods to study the real-time dynamics of generic quantum systems has been comparatively less prevalent~\cite{Mak:1992}. The dynamical sign problem may become more severe both with the size of the system, and duration for which it is simulated~\cite{Feynman:2012,Thirumalai:1983,Makri:1987}.  Despite these challenges, advances are being made in the treatment of these problems, including hybridization of Monte Carlo techniques with other methods~\cite{Makri:1988,Mak:1990,Makri:1991,Makri:1993,Jadhao:2008}.

The sign problem has been studied in the context of quantum computation, where it is known that a sufficient condition for efficient probabilistic classical simulation of the adiabatic evolution of a quantum system using Monte Carlo methods is that the Hamiltonian governing the quantum system is sign problem free (also known as stoquastic) and frustration free~\cite{Bravyi:2008,Bravyi:2009,Childs:2013}.  Projector Monte Carlo algorithms have been developed specifically for this type of problem~\cite{Bravyi:2008,Bravyi:2014}.  Moreover, the use of tools from quantum information allows any generic unitary evolution of a quantum system to be written as the ground state eigenvalue problem of a Hermitian Hamiltonian~\cite{Feynman:1982,Kitaev:2002,McClean:2013}.  In this work, we exploit this equivalence to adapt the interacting walker method introduced by Booth et. al~\cite{Booth:2009} to treat the dynamical sign problem with a method designed for the fermion sign problem.

The paper is organized as follows.  First, we review the time-embedded discrete variational principle~\cite{McClean:2013} and derive from it the Clock Hamiltonian~\cite{Feynman:1982,Kitaev:2002,McClean:2013}, which are the essential tools for writing a general unitary evolution as a ground state eigenvalue problem of a Hermitian Hamiltonian.  We then review the FCIQMC method and adapt it for application to the Clock Hamiltonian.  A discussion of the theoretical and practical manifestation of the dynamical sign problem in this setting follows with numerical examples from quantum computation.  Finally, we introduce a general framework of basis rotations which can be used to ameliorate the sign problem and study the performance of this method when used in parallel computation.

\section{Dynamics as a ground state problem}
It has been shown that any unitary quantum evolution may be formulated as a ground state eigenvalue problem with applications to classical simulation of quantum systems~\cite{McClean:2013}.  We briefly review the relevant results here so that this work remains self-contained.

Consider a quantum system that is described at discrete time steps $t$ by a normalized wavefunction $\ket{\Psi_t}$.  The dynamics of this system are described by a sequence of unitary operators $\{U_t\}$ such that
$U_t \ket{\Psi_t}=\ket{\Psi_{t+1}}$ and $U_{t}^{\dagger}\ket{\Psi_{t+1}} = \ket{\Psi_{t}}$.  From the properties of unitary evolution, the following is clear:
\begin{equation}
 2 - \bra{\Psi_{t+1}} U_t \ket{\Psi_t} - \bra{\Psi_t} U_t^\dagger \ket{\Psi_{t+1}} = 0.
\end{equation}

Moreover, if the wavefunctions at each point in time are only approximately known (but normalized) then
\begin{equation}
 \sum_t \left(2 - \bra{\Psi_{t+1}} U_t \ket{\Psi_t} - \bra{\Psi_t} U_t^\dagger \ket{\Psi_{t+1}} \right) \geq 0
\end{equation}
where equality holds only in the case where the wavefunction represents an exact, valid evolution of the quantum system.  To consider all moments in time simultaneously, we extend the physical Hilbert space with an ancillary quantum system to denote time.  This ancillary time register takes on integer values $t$ and is orthonormal such that $\braket{t'}{t} = \delta_{t,t'}$.  With this construction, we see that by defining
\begin{align}
\begin{split}
 \mathcal{H'} = \frac{1}{2} \Big( \sum_t \ & I \otimes \ket{t}\bra{t} - U_t \otimes \ket{t+1}\bra{t}  \\
   - & U_t^\dagger \otimes \ket{t}\bra{t+1} + I \ket{t+1}\bra{t+1} \Big)
\end{split}
\end{align}
which acts on the composite system-time Hilbert space, all valid time evolutions minimize
\begin{equation}
 \mathcal{S} = \sum_{t,t'} \bra{t'}\bra{\Psi_{t'}} \mathcal{H'} \ket{\Psi_t}\ket{t}.
\end{equation}
Note that we have adopted the convention of script letters for operators acting in the system-time Hilbert space such as $\mathcal{H'}$ as opposed to operators only acting on the system such as $U_t$. The time-embedded discrete variational principle immediately follows, which simply states that this quantity is stationary under variations of the wavefunction $\delta \ket{ \Psi_t}$ for all valid time evolutions, or
\begin{equation}
 \delta \mathcal{S} = \delta \sum_{t,t'} \bra{t'}\bra{\Psi_{t'}} \mathcal{H'} \ket{\Psi_t}\ket{t} = 0
\end{equation}

To select a particular evolution of interest, one may introduce a penalty operator that fixes the state of the system at a given time.  Typically, this might represent a known initial state, and this operator in the system-time Hilbert space is given by
\begin{equation}
 \mathcal{C}_0 = (I - \ket{\Psi_0}\bra{\Psi_0}) \otimes \ket{0}\bra{0}.
\end{equation}

The minimization of a Hermitian quadratic form constrained to have unit norm is equivalent to the eigenvalue problem for the corresponding Hamiltonian.  We introduce the Lagrange multiplier $\lambda$ to enforce normalization.  As both $\mathcal{S}$ and $C_0$ are Hermitian by construction, minimization of
\begin{align}
 \begin{split}
 \mathcal{L} = &\sum_{t,t'} \bra{t'}\bra{\Psi_{t'}} \mathcal{H'} + \mathcal{C}_0 \ket{\Psi_t}\ket{t} \\
  &- \lambda \left(\sum_{t,t'} \bra{t'}\braket{\Psi_{t'}}{\Psi_t}\ket{t} - 1 \right)
  \end{split}
\end{align}
is equivalent to solving for the eigenvector corresponding to the smallest eigenvalue of the Hermitian operator 
\begin{equation}
 \mathcal{H} = \mathcal{H'} + C_0
\end{equation}
which we refer to as the Clock Hamiltonian.  This Hamiltonian has a unique ground state with eigenvalue $0$ given by the history state,
\begin{equation}
 \ket{\Phi} = \frac{1}{\sqrt{T}} \sum_t \ket{\Psi_t}\otimes \ket{t}
\end{equation}
which encodes the entire evolution of the quantum system. Thus, the quantum dynamics of the physical system can be obtained by finding the ground state eigenvector of $\mathcal{H}$.

\section{FCIQMC for the Clock Hamiltonian}
The Full Configuration Interaction Quantum Monte Carlo (FCIQMC) method was introduced by Booth et. al as a method to accurately find the ground state for electronic structure problems in a basis of Slater determinants without appealing to the fixed node approximation to eliminate the fermion sign problem~\cite{Booth:2009}.  We review the basics of the theory behind this method and show how it can be adapted for the Clock Hamiltonian $\mathcal{H}$, such that it simulates the full time evolution of a quantum system.  

Let $\ket{\Phi_i}$ and $\lambda_i$ be the eigenvectors and corresponding eigenvalues of $\mathcal{H}$.  Any vector $\ket{\Psi}$ in the system-time Hilbert space acted upon by $\mathcal{H}$ can be decomposed in terms of the eigenvectors of $\mathcal{H}$ such that
\begin{equation}
 \ket{\Psi} = \sum_i c_i \ket{\Phi_i}
\end{equation}
It follows that for any $\ket{\Psi}$ not orthogonal to the ground state of the Clock Hamiltonian, $\ket{\Phi_0}$, that
\begin{equation}
 \lim_{\tau \rightarrow \infty} e^{-\tau \mathcal{H}} \ket{\Psi} = \lim_{\tau \rightarrow \infty} \sum_i e^{-\tau \lambda_i} c_i \ket{\Phi_i} \propto \ket{\Phi_0}
\end{equation}
Because $\mathcal{H}$ trivially commutes with itself, we may break this operator into the successive application of many operators, such that for a large number of slices $N$ of a finite $\tau$,
\begin{equation}
 e^{-\tau \mathcal{H}} = \left(e^{-\frac{\tau}{N} \mathcal{H}}\right)^N 
   \approx \left(1 - \delta \tau \mathcal{H}\right)^N
\end{equation}
where $\delta \tau = \tau/N$.  Note that the linearized time propagator used here is both simple to implement for discrete systems as well as unbiased in the final ($\tau \rightarrow \infty$) result  given some restrictions on $\delta \tau$~\cite{Trivedi:1990}.  Thus with a prescription to stochastically apply the operator 
\begin{equation}
 \mathcal{G} = \left(1 - \delta \tau \mathcal{H} \right)
\end{equation}
repeatedly to any vector in the system-time Hilbert space, we can simulate the quantum dynamics of the physical system.  $\tau$ is sometimes interpreted as imaginary-time by analogy to the Wick-rotated time-dependent Schr\"odinger equation, however we will only refer to $\tau$ as ``simulation time'' here, to avoid confusion with the simultaneous presence of real and imaginary-time.

To represent a vector in the system-time Hilbert space, we introduce discrete walkers represented by $\{i, t\}$ with associated real and imaginary integer weights $\mathcal{R}(\{i, t\})$ and $\mathcal{I}(\{i, t\})$.  These walkers correspond to a single system-time configuration.   The indices correspond to a system state $\ket{i}$ at time $t$ with a complex integer weight defined by its real and imaginary components, $W(\{k, t\}) = \mathcal{R}(\{k, t\}) + i \mathcal{I}(\{k, t\})$. Given a collection set of these walkers, the corresponding normalized vector is given by
\begin{equation}
 \ket{\Psi} = \frac{1}{K} \sum_{\{i, t\}}  W(\{i, t\}) \ket{i}\otimes\ket{t}
\end{equation}
where $K$ is the normalization constant given by the sum of the absolute value of all the complex integer walker weights
\begin{align}
K = \sum_{\{i, t\}} |W(\{i, t\})|.
\end{align}
We also use this notation to define matrix elements for an operator $\mathcal{O}$ between a state $\ket{i}\ket{t}$ and $\ket{j}\ket{t'}$ as
\begin{align}
\mathcal{O}_{\{j,t'\},\{i,t\}} = \bra{j}\bra{t'} \mathcal{O} \ket{i} \ket{t}
\end{align}

To stochastically apply the operator $\mathcal{G}$ to a vector represented by a set of such walkers, the following four steps are used, adapted from the original implementation by Booth et. al:
\begin{enumerate}
  \item Spawning: This step addresses the off-diagonal elements of $\mathcal{G}$.  For each walker $\{i,t\}$, we consider $N_r = \mathcal{R}(\{i, t\})$ real parents and $N_i = \mathcal{I}(\{i, t\})$ imaginary parents, both with the correct associated sign.  For each of the real parents $N_i$, we select a coupled state at an adjacent time and attempt to spawn a real child and imaginary child $\{j,t'\}$ with probabilities
  \begin{align}
   p_s^\mathcal{R}(\{j,t'\}|\{i,t\}) &= \frac{\delta \tau \left|\mathcal{R}(\mathcal{H}_{\{j,t'\},\{i,t\}})\right|}{p_{g_t}(t',t)p_{g_s}(\{j,t'\}|\{i,t\})} \\
   p_s^\mathcal{I}(\{j,t'\}|\{i,t\}) &= \frac{\delta \tau \left|\mathcal{I}(\mathcal{H}_{\{j,t'\},\{i,t\}})\right|}{p_{g_t}(t',t)p_{g_s}(\{j,t'\}|\{i,t\})} 
  \end{align}
  with corresponding signs
  \begin{align}
   S^\mathcal{R} &= -\text{sign}(\mathcal{R}(\mathcal{H}_{\{j,t'\},\{i,t\}})) \\
   S^\mathcal{I} &= -\text{sign}(\mathcal{I}(\mathcal{H}_{\{j,t'\},\{i,t\}}))
  \end{align}

  and for each of the imaginary parents $N_i$ we select a coupled state at an adjacent time and attempt to spawn a real child and imaginary child $\{j,t'\}$ with probabilities
  \begin{align}
   p_s^\mathcal{R}(\{j,t'\}|\{i,t\}) &= \frac{\delta \tau \left|\mathcal{I}(\mathcal{H}_{\{j,t'\},\{i,t\}})\right|}{p_{g_t}(t',t)p_{g_s}(\{j,t'\}|\{i,t\})} \\
   p_s^\mathcal{I}(\{j,t'\}|\{i,t\}) &= \frac{\delta \tau \left|\mathcal{R}(\mathcal{H}_{\{j,t'\},\{i,t\}})\right|}{p_{g_t}(t',t)p_{g_s}(\{j,t'\}|\{i,t\})}
  \end{align}
  with corresponding signs
  \begin{align}
   S^\mathcal{R} &= \ \ \text{sign}(\mathcal{I}(\mathcal{H}_{\{j,t'\},\{i,t\}})) \\
   S^\mathcal{I} &= -\text{sign}(\mathcal{R}(\mathcal{H}_{\{j,t'\},\{i,t\}}))
  \end{align}

  where probabilities $p_s > 1$ are handling by deterministically spawning $\floor{p_s}$ walkers and spawning an additional walker with probability $p_s - \floor{p_s}$. $\delta \tau$ is the simulation time step and may be used to control the rate of walker spawning.  The functions $p_{g_t}(t',t)$ and $p_{g_s}(\{j,t'\}|\{i,t\})$ are the probability of suggesting a walker at the new time $t'$ and of the particular state $j$ respectively.  For the Clock, an efficient choice of the time generation function, $p_{g_t}(t',t)$ is $t'=t \pm 1$ with equal probability unless the walker is at a time boundary, in which case it should move inward with unit probability.  The state generation probability $p_{g_s}(\{j,t'\}|\{i,t\})$ should be chosen based on knowledge of the structure of $U_t$ such that connected states may reach each other.  In this work we use a uniform distribution where states connected by $U_t$ are selected randomly with equal probability, however this can be refined using knowledge of $U_t$.

  In this case, where $\{j,t'\}\neq\{i,t\}$, the matrix elements $\mathcal{H}_{\{j,t'\},\{i,t\}}$ may be written more explicitly as
  \begin{equation}
   \mathcal{H}_{\{j,t'\},\{i,t\}} = 
    \begin{cases}
      -\frac{1}{2}\bra{j}U_t\ket{i} & t'=t+1 \\
      -\frac{1}{2}\bra{j}U_t^\dagger\ket{i} & t'=t-1 \\
      0 & \text{otherwise}
    \end{cases}
  \end{equation}

  \item Diagonal Death/Cloning: This step addresses the application of the diagonal of $\mathcal{G}$.  In this step, for each parent walker $\{i,t\}$ (not yet including child walkers spawned in the last step), calculate
  \begin{equation}
   p_d(\{i,t\}) =  \delta \tau (\mathcal{H}_{\{i,t\},\{i,t\}} - S)
  \end{equation}
  where $S$ is a shift that is used to control the population in the simulation.  Now for each real and imaginary parent $N_r$ and $N_i$ associated with $\{i,t\}$,  if $p_d > 0$, the parent is destroyed.  If $p_d<0$, the parent is cloned with a probability $|p_d|$, handling instances of greater than unit probabilities as in the previous step.

  In the case of the Clock, the diagonal matrix elements take on a simple form
  \begin{equation}
    \mathcal{H}_{\{i,t\},\{i,t\}} = 
    \begin{cases}
     1/2 + (1 - |\braket{i}{\Psi_0}|^2) & t = 0 \\
     1/2 & t = T-1 \\
     1 & \text{otherwise}
    \end{cases}
  \end{equation}

  \item Annihilation:  In this step, all previously existing and newly spawned walkers are searched, and any which match are combined such that both their real and imaginary components are summed.  In the event that any walker ends up with $0$ total weight, it is removed entirely from the simulation.  In the case of the Clock, it is advantageous to store walkers grouped by time, such that in parallel implementations the simulation can be easily split along this dimension. This will be elaborated upon later.  Within each group it is advisable to use any natural ordering present on the basis states to enable binary search that can locate identical walkers in a time that is logarithmic in the number of walkers at a given time.  Alternatively one can use hash tables to facilitate annihilation~\cite{Booth:2013A}.
  
\end{enumerate}

\begin{figure}
\includegraphics[width=8.6 cm]{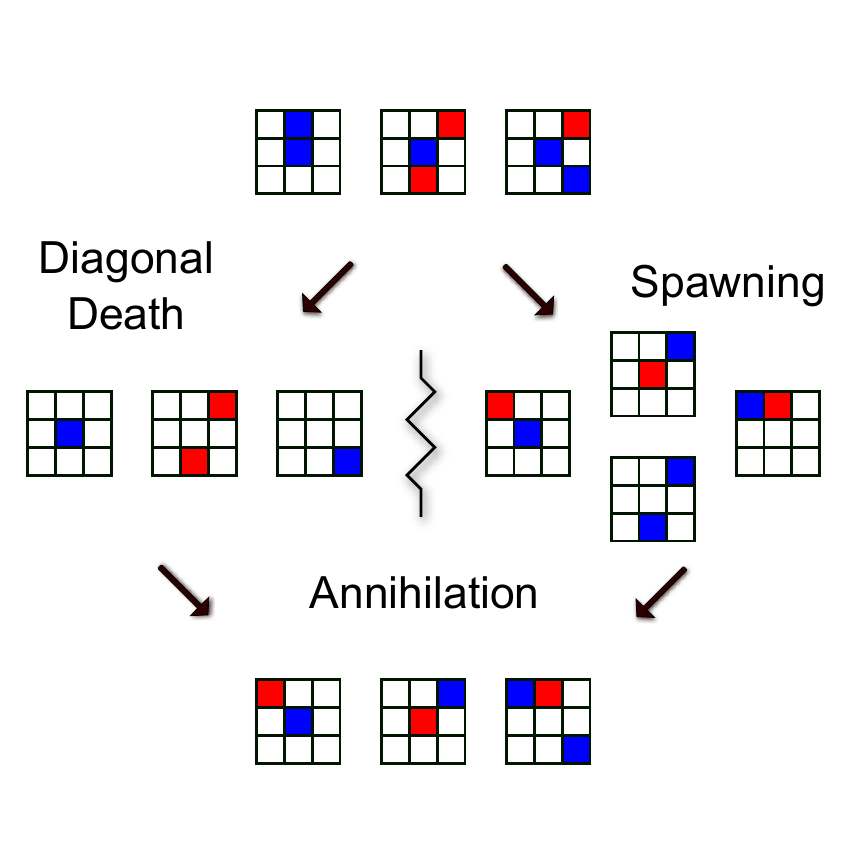}
\caption{A schematic representation of a single iteration of the FCIQMC algorithm for the Clock.  The larger squares represent real-time, and sub-squares represent the possible quantum states occupied by either positive (blue) or negative (red) walkers.  In each iteration, the set of parents spawns potential children to adjacent times, with parentage being indicated by dotted lines.  Simultaneously the set of parent walkers are considered for diagonal death.  Finally, the remaining set of parents and spawned children are combined, allowing walkers with opposing signs at the same state and time to annihilate. }
\label{fig:FCIQMCOutline}
\end{figure}

 A single iteration of the above algorithm is cartooned in Fig. \ref{fig:FCIQMCOutline}. By using this procedure, the operator $G$ is iteratively applied until the state of walkers is equilibrated at some simulation time $\tau > \tau_{eq}$, with a number of walkers $N_w$.  The average of some observable $O$ may be estimated at simulation time $\tau$ according to
 \begin{equation}
  \avg{O}(\tau) = \frac{\bra{\Phi(\tau)}O \ket{\Phi(\tau)}}{\braket{\Phi(\tau)}{\Phi(\tau)}}
 \end{equation}
 By averaging over the simulation time $\tau$ and correcting for the autocorrelation time of the quantity $\avg{O}$ using standard statistical procedures, the average may be converged to the desired precision.  In general, however, the simulation time averaged quantity $\avg{O}_{\tau}$ may be biased due to the sign problem~\cite{Spencer:2012,Kolodrubetz:2013,Shepherd:2014}.  This bias may be removed to an arbitrary degree by increasing the number of walkers $N_w$ such that the state remains sign-coherent between steps.  The number of walkers required to remove the bias to a given precision depends both on the severity of the sign problem and the amount of Hilbert space the physical problem occupies~\cite{Spencer:2012,Kolodrubetz:2013,Shepherd:2014}.  To this end, we define a problem-dependent number $n_c$ such that when $N_w > n_c$, the time averaged quantity $\avg{O}_{\tau}$ is accurate to the desired precision.  Because this is an NP-Complete problem, one must expect that in general, $n_c$ is on the order of the dimension of the Hilbert space, $D$, that is, it grows exponentially with the size of the system and linearly with real-time.  However, for many systems of interest in ground state problems it has been found that $n_c << D$~\cite{Booth:2010,Booth:2013,Shepherd:2014}, and one might expect the same to be true for some dynamical problems.  We now turn our attention to the scaling and properties of $n_c$ for dynamical systems.

\section{Manifestation of the sign problem}
The conditions for the efficient simulation of a Hamiltonian on a classical computer have been studied in the context of quantum complexity theory.  It is known that if a Hamiltonian is frustration free and has real, non-positive off-diagonal elements in a standard basis (stoquastic) that it may be probabilistically simulated to a set precision in a time that is polynomial in the size of the system~\cite{Bravyi:2008,Bravyi:2009}.  

For practical purposes, there are limitations on the system operators one may simulate.  In particular, the system operators must be the sum of a polynomial number of terms.  This simply originates from the need to be able to efficiently evaluate matrix elements of a given state.  The interaction of at most $k$ particles, or $k-$local interactions, in the physical system is a sufficient but not necessary condition for this to be true.  The Clock construction has also been recently generalized to open quantum systems~\cite{Tempel:2014}, where even in this case a $2-$local construction is generally possible with the use of gadgets.  Alternatively, if the Clock is constructed from a sequence of unitary  gates that act on at most $k$ qubits in quantum computation, then the Clock will also satisfy this requirement.  

The presence of frustration in interacting systems can cause the autocorrelation time of measured observables to diverge exponentially, rendering their efficient simulation intractable even in cases where the Hamiltonian is bosonic or sign problem free~\cite{Troyer:2005,Wei:2010}. It has been proven generally that the Clock Hamiltonian is frustration free, with a unique ground state separated from the first excited state with a gap of $O(1/T^2)$ where $T$ is the number of discrete time steps being considered at once. 

If an operator is stoquastic (or sign problem free), then the off-diagonal elements that correspond to transitions in a Monte Carlo simulation all be non-positive.  The operator $\mathcal{G}$ will contain only positive transition probabilities in this case and have a ground state corresponding to a classical probability distribution by the Perron-Frobenius theorem~\cite{Bravyi:2008,Bravyi:2009}.  In the context of the FCIQMC method introduced, this means that walkers will never change signs throughout the simulation, and all averages will be sign-coherent and unbiased independent of the number of walkers $N_w$. In the Clock Hamiltonian, the off-diagonal elements correspond to the set of unitary operators with their adjoints $\{U_t,U^\dagger_t\}$, and the penalty term $\mathcal{C}_0$.  For the standard computational initial state ($\ket{0}^{\otimes N}$), the penalty term $\mathcal{C}_0$ has a fixed sign, and thus the Clock Hamiltonian is stoquastic if $\{U_t,U^\dagger_t\}$ represented in the standard basis has all real positive entries, yielding non-positive off-diagonal entries in the Clock.

Given the ubiquity of $k-$local interactions in physical problems and the rigorous proof that the Clock Hamiltonian is frustration free, we will take these two conditions as given and consider more carefully the stoquastic condition.  Consider a simple example of a unitary evolution that may be simulated on a classical computer efficiently, namely reversible classical computational.  All reversible classical circuits may be expressed in terms of Toffoli gate sequences, which is unitary and acts to switch a target bit conditional on the state of two control bits.  In the standard computational basis it has a representation given by
\begin{equation}
Tof = \left( \begin{array}{cccccccc}
1 & 0 & 0 & 0 & 0 & 0 & 0 & 0 \\
0 & 1 & 0 & 0 & 0 & 0 & 0 & 0 \\
0 & 0 & 1 & 0 & 0 & 0 & 0 & 0 \\
0 & 0 & 0 & 1 & 0 & 0 & 0 & 0 \\
0 & 0 & 0 & 0 & 1 & 0 & 0 & 0 \\
0 & 0 & 0 & 0 & 0 & 1 & 0 & 0 \\
0 & 0 & 0 & 0 & 0 & 0 & 0 & 1 \\
0 & 0 & 0 & 0 & 0 & 0 & 1 & 0
\end{array}
\right)
\end{equation}
The Clock when constructed with unitary Toffoli gates is stoquastic and $n_c \approx 1$.  Although a stoquastic Hamiltonian is sufficient for this to be the case, it is not a necessary condition.  To see this, consider a slightly different set of unitary operators, namely the standard Pauli group gates, $X_i$, $Y_i$, and $Z_i$ in combination with the CNOT gate.  These gates have the following unitary representations in the standard computational basis
\begin{align}
X &= \left( \begin{array}{cc}
0 & 1 \\
1 & 0 \end{array} \right) \\
Y &= \left( \begin{array}{cc}
0 & -i \\
i & 0 \end{array} \right) \\
Z &= \left( \begin{array}{cc}
1 & 0 \\
0 & -1 \end{array} \right) \\
CNOT &= \left( \begin{array}{cccc}
1 & 0 & 0 & 0 \\
0 & 1 & 0 & 0 \\
0 & 0 & 0 & 1 \\
0 & 0 & 1 & 0 \end{array} \right)
\end{align}
Considering for now only the computational basis we simulate in (a restriction we later lift), it is clear that given the complex entries and varying signs of the off-diagonals, that a Clock Hamiltonian built from this gate set will not be stoquastic if even a single $Y$ or $Z$ gate is used.  However these gates also have the property that they map single configurations to single configurations, and as as a result no interference occurs, yielding all sign coherent averages and $n_c \approx 1$. We call this type of transformation, which is configuration number preserving, ``quasi-classical'', in contrast to classical which we define as configuration number preserving as well as phase preserving. Thus a stoquastic Clock Hamiltonian is a sufficient, but not necessary condition for the simulation procedure to be sign-problem free.

Consider a slightly more general local rotation $R$ parameterized by an angle $\theta$ 
\begin{equation}
  R(\theta) = \left( \begin{array}{cc}
  \cos \theta  & \sin \theta \\
  -\sin \theta & \cos \theta \end{array} \right)
\end{equation}
In this case, the value of $n_c$ as a function of system size is more complex.  These represent the real-time evolutions of local spin Hamiltonians for systems of spin-$\frac{1}{2}$ particles.

\begin{figure}[ht]
\centering
\includegraphics[width=8cm]{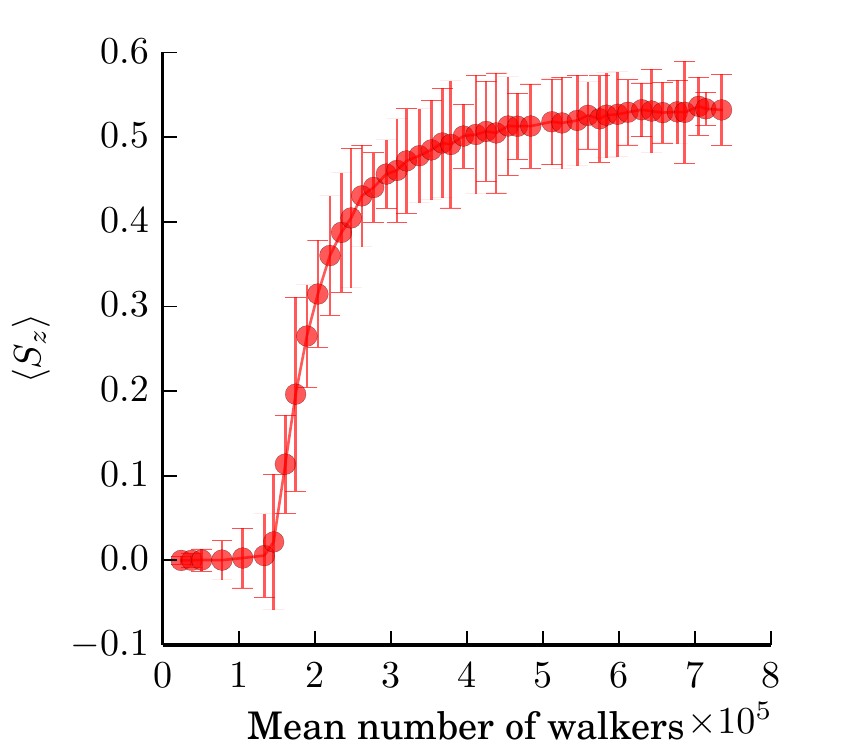}
\caption{Computed expectation value for $S_z$ for a single qubit at the final time in the simulation as a function of the average number of walkers kept in the simulation.  There are 11 total qubits in the simulation.  It is apparent the system exhibits a sharp transition between a totally sign incoherent sampling where all averages become zero, and a sign coherent region where the averages begin to converge to the appropriate value.}
\label{fig:singleRot}
\end{figure}

In Fig. \ref{fig:singleRot} we consider a single rotation $R(\theta)$ with $\theta=5\pi/32$ applied uniformly to 11 qubits initialized to $\ket{0}^{\otimes N}$.  As the Clock Hamiltonian in this simulation is neither stoquastic nor quasi-classical, one observes a sign-incoherent region for a small number of walkers, where all averages tend towards $0$, until some critical threshold $N_w > n_c$ is reached where a transition occurs to sign-coherent sampling, and the average converges to the true value.  We note that some implementations of the FCIQMC algorithm have used the diagonal shift $S$ as a proxy for convergence~\cite{Booth:2009}, but we did not observe a similar plateau trend here.  The history state being sampled in this case is given by
\begin{align}
\ket{\Psi} = \sum_t \frac{1}{\sqrt{T}} \left( R(\theta) \ket{0} \right)^{\otimes t} \ket{0}^{\otimes T-t} 
\end{align}

\begin{figure}[ht]
\centering
\includegraphics[width=8cm]{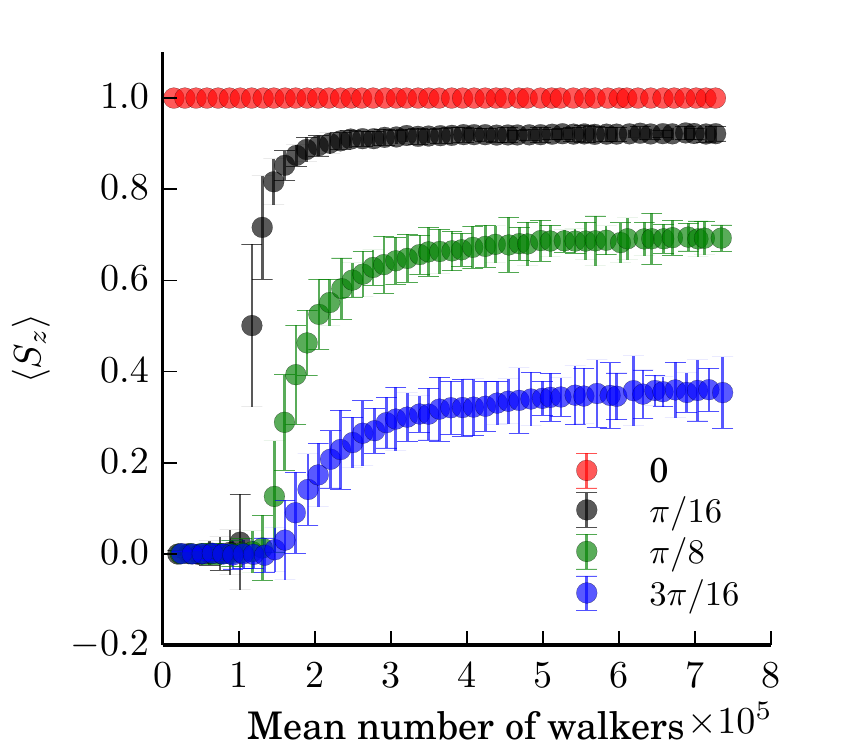}
\caption{Computed expectation value for $S_z$ for a single qubit at the final time in the simulation as a function of the average number of walkers kept in the simulation and the rotation angle used in the simulation.  The rotation angle $\theta$ is indicated by the line label.  The simulation contains 11 total qubits for all rotation angles.  One sees that the closer the rotation is to quasi-classical, the sharper and earlier the transition to sign coherent sampling.}
\label{fig:manyRot}
\end{figure}

The formal structure of this evolution is quite similar for all values of $\theta$, however the states that result can exhibit quite different features with respect to the sign problem in sampling.  In Fig. \ref{fig:manyRot} we examine the same circuit, but include many different rotation angles.  One sees that not only does the value $n_c$ change as a function of rotation angle, but the rate of the transition is quite different as well, favoring sharper, earlier transitions for rotations that are closer to quasi-classical ($\theta=0$).

\section{Mitigating the sign problem}
In the last section we considered the impact of sign problem as it related to local rotations  (or the dynamics of distinguishable non-interacting particles).  The apparent challenges in this domain are unsettling given that trivial solutions are known for this problem.  Here we propose a simple scheme to mitigate the sign problem to an arbitrary extent using preliminary computation.

It is known that the sign problem is generically basis dependent.  To this end we propose an analogous approach to the interaction picture in quantum dynamics, where the walkers at each point in time are expressed in a new basis determined by a generic time-dependent unitary rotation given by $\{\mathcal{B}_t\}$.  The evolution operators are also dressed by this change such that in the new basis, the Clock is constructed from the rotated operators given by
\begin{equation}
 U'_t = \mathcal{B}_{t+1}^\dagger U_t \mathcal{B}_{t}
\end{equation}

Moreover, the computation of any Hermitian observable $O$ must also take into consideration the new basis, such that 
 \begin{equation}
  \avg{O}(\tau) = \frac{\bra{\Phi(\tau)} \mathcal{B}_t^\dagger O \mathcal{B}_{t} \ket{\Phi(\tau)}}{\braket{\Phi(\tau)}{\Phi(\tau)}}
 \end{equation}

If one finds a set of $\{\mathcal{B}_t\}$ that renders the Clock Hamiltonian stoquastic or quasi-classical, the resulting Hamiltonian may be sampled readily.   One expects that in general, finding this basis must be at least as difficult as solving exactly the problem of the quantum evolution.  In fact, it is easy to see that one may choose the exact evolution as the set of basis rotations, and that this renders the Clock Hamiltonian stoquastic and trivial, such that the evolution is dictated by the identity at all times.  Of course the price one must pay for this is that the computation of observables requires the full evolution to be known.

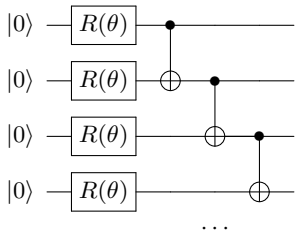
\begin{figure}
\centering
$\Qcircuit @C=1em @R=.7em {
 \lstick{\ket{0}} & \gate{R(\theta)} & \ctrl{1} & \qw & \qw & \qw\\
 \lstick{\ket{0}} & \gate{R(\theta)} & \targ & \ctrl{1} & \qw & \qw\\
 \lstick{\ket{0}} & \gate{R(\theta)} & \qw & \targ & \ctrl{1} \qw & \qw\\
 \lstick{\ket{0}} & \gate{R(\theta)} & \qw & \qw & \targ & \qw \\
 & & & \cdots
 }$
\caption{The quantum circuit diagram for the circuit used to test the efficacy of time-dependent local rotations in ameliorating the sign problem.  The angle used in this case is $\theta=0.49$.  We compare the results from this circuit as a function of the number of controls that are removed from the NOT gates (crossed circle here), and whether time dependent local basis rotations are utilized.  The controls are removed from the end of the circuit first.}
\label{fig:RotCircuit}
\end{figure}

However, as was seen above, it is not necessary for the Clock to be rendered completely trivial. Even approaching a quasi-classical Hamiltonian in an approximate sense can greatly reduce the sampling costs.  For some instances, one may find an approximate set of rotations that make the Clock nearly stoquastic or quasi-classical, and the remainder of the sign problem can be handled by maintaining a reasonable number of walkers $N_w$ in the simulation.  As an example of this procedure, we consider the simple case where $\{\mathcal{B}_t\}$ are determined entirely by the local rotations in a quantum circuit.  Specifically, for local rotations, $\mathcal{B}_t = \prod_{t' < t}^{0} U_{t'}$, where $U_{t'}$ is replaced by $I$ for two- or more qubit operations.  It is clear that for circuits consisting of only local rotations, as in the previous section, this is equivalent to exact evolution and the resulting Clock Hamiltonian becomes trivial ($U_t' = I \ \forall \ t'$).  To study how this works in the non-trivial case, we examine a similar circuit of local rotations, but now with a variable number of CNOT gates included.  This elucidates to what extent the use of basis rotations can mitigate the sign problem when they are not an exact solution to the dynamics considered.  A depiction of this circuit is given in Fig. \ref{fig:RotCircuit}.

\begin{figure}[ht]
\centering
\includegraphics[width=8cm]{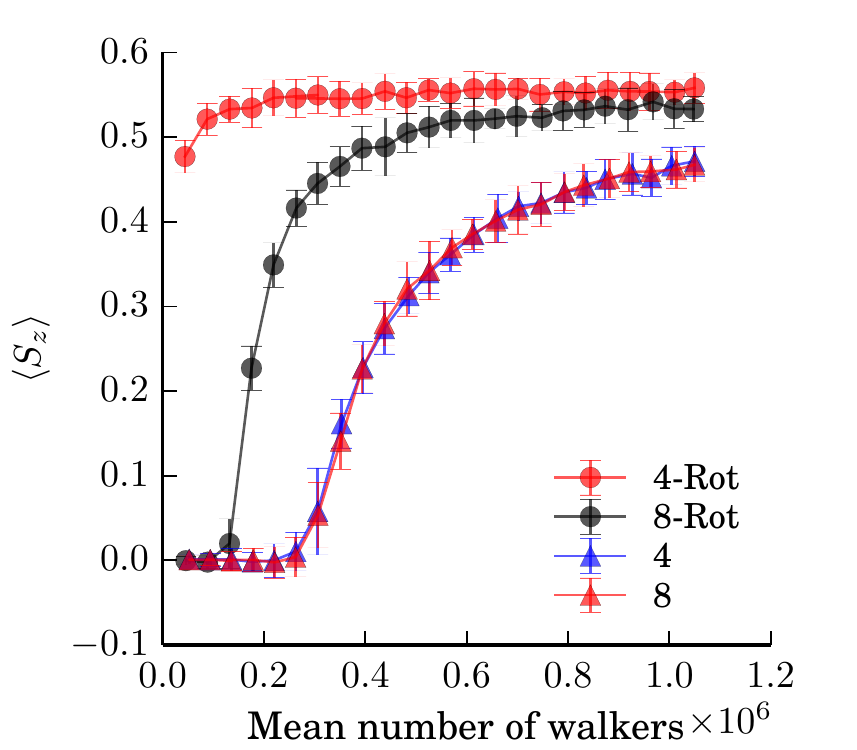}
\caption{The mean value of a spin observable is plotted as a function of the mean number of walkers labeled by the number of CNOT gates both with local basis rotation(-Rot) and without.  It is seen that even for a relatively high proportion of CNOT gates, the rotated basis performs far better than the non-rotated basis, requiring a lower number of walkers to reach sign-coherent sampling.}
\label{fig:basisRot}
\end{figure}

In Fig. \ref{fig:basisRot} we consider a simple quantum circuit consisting of a series of local rotations followed by NOT gates, with a variable number of the NOT gates under control. With the given rotation angle ($\theta = 0.49$), these are entangling operations not covered by the simple local basis rotation scheme we use here. However, it is seen that even for 8 controlled NOT gates, the use of local basis rotations dramatically reduces the number of walkers required to reach sign-coherent sampling, indicating this scheme can be computationally effective even for simulations containing a considerable fraction of two-qubit entangling operations.  Further tests of more complex quantum circuits are needed to determine the efficiency of different rotation schemes as a function of the structure of the quantum circuit.

\section{Parallel-in-time scaling}
Monte Carlo methods are often championed as the ultimate parallel algorithms, associated with the phrase ``embarrassingly parallel''.  Given the evolution of modern computational architectures towards many-core architectures with slower Clock speeds, Monte Carlo will continue to play a growing role in the numerical simulation of physics at the boundaries of our computational capabilities.  Interacting walker Monte Carlo methods, can be more difficult to parallelize effectively due to the annihilation step where communication of walkers is unavoidable.  

\begin{figure}[ht]
\includegraphics[width=8.6 cm]{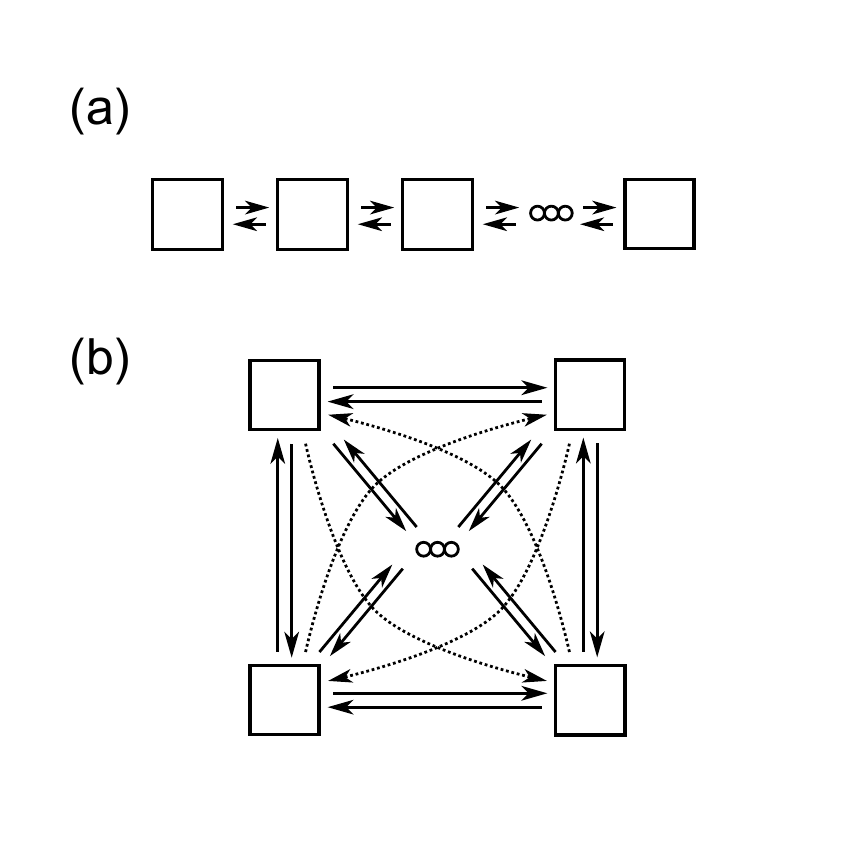}
\caption{A schematic representation of the communication patterns the annihilation step of interacting walker Monte Carlo schemes, where the boxes represent different MPI processes and the ellipsis represents the rest of the processes. In the case of the Clock (a), a time domain decomposition allows one to restrict communication to only nearest neighbour processes, facilitating simple, constant time communication amenable to the architecture of modern parallel computers.  In the more general case (b), a clear partitioning may not be readily achievable, and all processes may need to communicate with all other processes, creating a bottleneck.}
\label{fig:CommDiagram}
\end{figure}

In contrast to the most general interacting walker algorithm, which may require heavy communication between all processes, the FCIQMC method applied to the Clock Hamiltonian may take advantage of time-locality to create an efficient parallel-in-time algorithm using the standard method of domain decomposition in time.  Using this construction, only processes containing adjacent times need to communicate their child walkers, which may be done simultaneously in a time that is constant for the number of processes involved.  This remains true so long as the number of time steps under consideration is larger than the number of processes in use, which is typically the case.  In the case that the number of processes is much greater than the number of timesteps, this scheme may still be used by blocking multiple processors to each time, and utilizing an all-to-all communication pattern within each block only.  The difference between these two communication patterns is highlighted in Fig. \ref{fig:CommDiagram}.

\begin{figure}[ht]
\centering
\includegraphics[width=8cm]{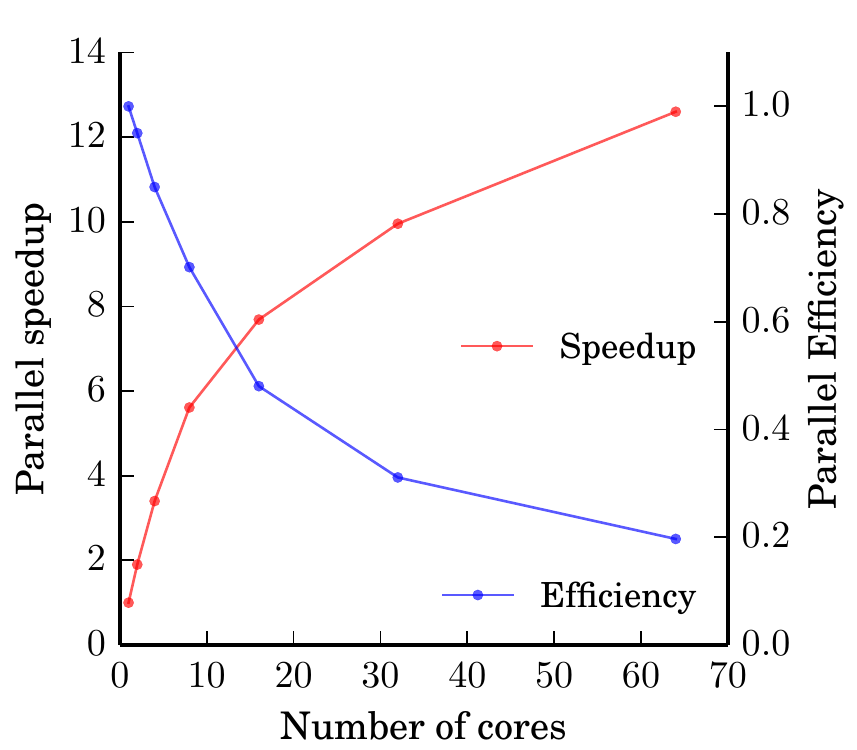}
\caption{A scaling study of our method implementation with a fixed total problem size (strong scaling), showing parallel efficiencies and speedups. The simulation consisted of $11$ qubits with $128$ time points generated by consecutive local rotations with $\theta \approx 0.098$.  The simulation maintained on average $10^6$ walkers in each simulation-time step and the wall clock time was measured to the point of an equivalent number of statistical samples.}
\label{fig:strongScaling}
\end{figure}

To demonstrate the scaling properties of this approach, we consider the scaling as a function of the number of processors for fixed total problem size, or strong scaling, with our implementation.  This benchmarking is done on a standard Linux cluster composed of AMD Opteron 6376 processors.  The parallel speedup with respect to single core time as a function of the number of processors is given in Fig \ref{fig:strongScaling}.  Here, we see that we are able to combine the parallelism of Monte Carlo with the locality of time-decomposition to achieve practical parallel efficiencies of over $95$\% with $2$ processors and $70$\% with $8$ using a simple MPI implementation on a commodity cluster.

\section{Conclusions}
In this work we reviewed the mapping between unitary dynamics and ground state eigenvalue problems.  We then showed how the FCIQMC method, a technique originally designed to ameliorate the fermionic sign problem for ground state electronic systems, could be applied to quantum dynamics problems as a direct result of this mapping.   This establishes a potential research direction for explicit connections between the fermionic and dynamical sign problems that plague quantum Monte Carlo simulations, and provides a pathway for the transfer of tools between the two domains.  

The numerical consequences of the dynamical sign problem in this context were studied using a few basic quantum circuits.  It was found that even local rotations can exhibit a severe sign problem depending on the form of the rotation and how different it is from a quasi-classical operation.  We then introduced a general method analogous to the interaction picture in dynamics or natural orbitals in the study of eigenstates that uses basis rotations to mitigate the difficulty of the problem.  The costs and benefits of different types of rotations require further research, however we showed that even local rotations can have a significant benefit for non-trivial circuits. Finally, we discussed the structure of the problem in the context of parallel-in-time dynamics, and showed high parallel efficiencies with only a basic MPI implementation on a commodity cluster.

Overall, we believe this is a promising new method for the simulation of quantum dynamics. It clarifies the bridge between dynamics and ground state problems and is capable of effectively utilizing parallel computing architectures.  While we have only demonstrated it for quantum circuits, we believe it will be generally useful for the study of quantum dynamics.

\section{Acknowledgements}
We acknowledge Ryan Babbush and Thomas Markovich for their helpful comments on the manuscript.  J.M. is supported by the Department of Energy Computational Science Graduate Fellowship under grant number DE-FG02-97ER25308. A.A-G. acknowledges the support of the Air Force Office of Scientific Research under Award No. FA9550-12-1-0046 and the National Science Foundation under Award No. CHE-1152291.

\bibliographystyle{apsrev4-1}
\bibliography{ClockMC}
\end{document}